\documentclass[onecolumn,preprintnumbers,amsmath,amssymb]{revtex4}

\usepackage{graphicx}
\usepackage{amssymb}
\usepackage{mathrsfs}
\usepackage{bm}
\usepackage{color}
\usepackage{amsmath}
\usepackage{mathrsfs}
\usepackage{algorithm}
\usepackage{algorithmic}
\usepackage{mathtools}
\usepackage{graphicx}
\usepackage{subfigure}
\usepackage{float}
\usepackage{threeparttable}

\usepackage{amsmath}
\usepackage{blindtext}
\usepackage{amsmath}
\usepackage{bbm}
\usepackage{float}
\usepackage{subfigure}
\usepackage{url}
\usepackage{multirow}

\newcommand{\sA}{\textsf{A}}
\newcommand{\sB}{\textsf{B}}
\newcommand{\sC}{\textsf{C}}

\newcommand{\cB}{\mathcal{B}}

\newcommand{\cH}{\mathcal{H}}

\newcommand{\cG}{\mathcal{G}}
\newcommand{\be}{\begin{eqnarray}}
\newcommand{\ee}{\end{eqnarray}}

\newcommand{\ba}{\begin{array}}
\newcommand{\ea}{\end{array}}

\begin{document}

\title{Verification of Bell Nonlocality by Violating Quantum Monogamy Relations}

\author{Yan-Han Yang$^{1}$, Xin-Zhu Liu$^{1}$,Xing-Zhou Zheng$^1$,Shao-Ming Fei$^{2,3}$, Ming-Xing Luo$^{1,4}$}

\affiliation{$^{1}$ School of Information Science and Technology, Southwest Jiaotong University, Chengdu 610031, China
\\
$^2$ School of Mathematical Sciences, Capital Normal University, Beijing 100048, China
\\
$^3$ Max-Planck-Institute for Mathematics in the Sciences, 04103 Leipzig, Germany
\\
$^{4}$ CAS Center for Excellence in Quantum Information and Quantum Physics, Hefei, 230026, China
}

\begin{abstract}
Quantum nonlocality as a witness of entanglement plays a crucial role in various fields. Existing quantum monogamy relations rule out the possibility of simultaneous violations of any Bell inequalities with partial statistics generated from one Bell experiment on any multipartite entanglement or post-quantum sources. In this paper, we report an efficient method to construct multipartite Bell test based on any Bell inequalities. We demonstrate that violating these monogamy relations can dynamically witness simultaneous Bell nonlocalities of partial systems. We conduct a tripartite experiment to verify quantum nonlocalities by violating a tripartite monogamy relation using a maximally entangled two-photon state.

\textbf{Keywords}: Quantum nonlocality; multipartite entanglement; quantum monogamy relations; dynamical witness; entangled two-photon

\end{abstract}

\maketitle


\section*{Introduction}

 Quantum entanglement is a fascinating phenomenon that has attracted significant attention in the field of quantum mechanics. Local measurements on bipartite entangled systems can give rise to joint correlations that exhibit nonlocal features, which cannot be explained by classical physics under the assumptions of locality and causality \cite{EPR,Bell}. These bipartite nonlocal correlations can be witnessed using Bell inequalities \cite{CHSH,Li,Son,BCPS}, which provide device-independent methods for verifying quantum nonlocality \cite{CHSH,BCPS}. In multipartite scenarios, there are richer nonlocal correlations. one example is a simple extension of bipartite nonlocal correlations to characterize the bipartition of all the systems. The other is the genuine multipartite correlations  \cite{Sy} that can witness the most strongest global nonlocal correlations generated from the multipartite entanglement to rule the classical mixture of any biseparable correlations. This has been extended for  GHZ states \cite{Aolita2012} and graph states \cite{Guhne2005}, operational frameworks \cite{Gallego2012}, network entanglement \cite{Luo2021a,Armin2022} or robusted entanglement \cite{Luo2021b}. Quantum entanglement has potential applications in quantum computing \cite{Raussendorf,Linden,Zurel,McArdle,Hangleiter}, quantum cryptography \cite{Curty,Lo,Ma,Vazirani,Yin,Zhang2,Gisin,Xu,Portmann}, and communication \cite{ZZH,HHH,Kim,Bennett,Cavalcanti,Lee,Hu}.

Witnessing the quantum nonlocality of entanglement can be equivalently realized by testing the winning probability of specific nonlocal games \cite{BCPS,SMA}. A nonlocal game is a hypothetical game in which several players, each receive a question from a referee, and then respond with an answer. The referee randomly selects the questions according to a known distribution, and, upon receiving answers from all players, decides whether they win or lose. The violation of a Bell inequality indicates that quantum players using quantum resources have a higher probability of winning in an equivalent nonlocal game than those using classical resources. One example is the Clauser-Horne-Shimony-Holt (CHSH) game \cite{CHSH} for evaluating the Boolean equation: $a\oplus b=x\wedge y$. This has been extended to witness nonlocality of general entanglement \cite{RV,EWL,Luo2019}.

All the previous Bell tests have assumed that the observers share a classical variable or quantum state. This setting in principle makes use of static implementations of Bell experiments, i.e., the source generates the fixed states and then distributes to observers. This is reasonable for standard statistic assumption of independent and identical distributions. But for multipartite quantum entanglement, it exhibits local monogamy features \cite{Coffman,Terhal,Osborne2006}, which limit the sharing of local entanglement between different subsets of particles. For example, although two particles can be entangled with each other, but they cannot simultaneously share the same level of entanglement with another particle in a multipartite entangled system. This implies that the bipartite nonlocality verified in local Bell test will rule out the possibilities of all the other bipartite nonlocal correlations even if they may jointly exhibit the genuine multipartite nonlocality. This can be formally illustrated using a tailed CHSH inequality for each pair in a multipartite Bell test, where any quantum or post-quantum resources show no advantage beyond classical resources \cite{Toner2009,RH,Regula2014,PB,Fei,LiF}. This raises a question of how the monogamous nature of global multipartite entanglement affects Bell tests of bipartite nonlocalities \cite{Kurzy,Tran}.

\begin{figure}[h!]
\begin{center}
\includegraphics[scale = 0.8]{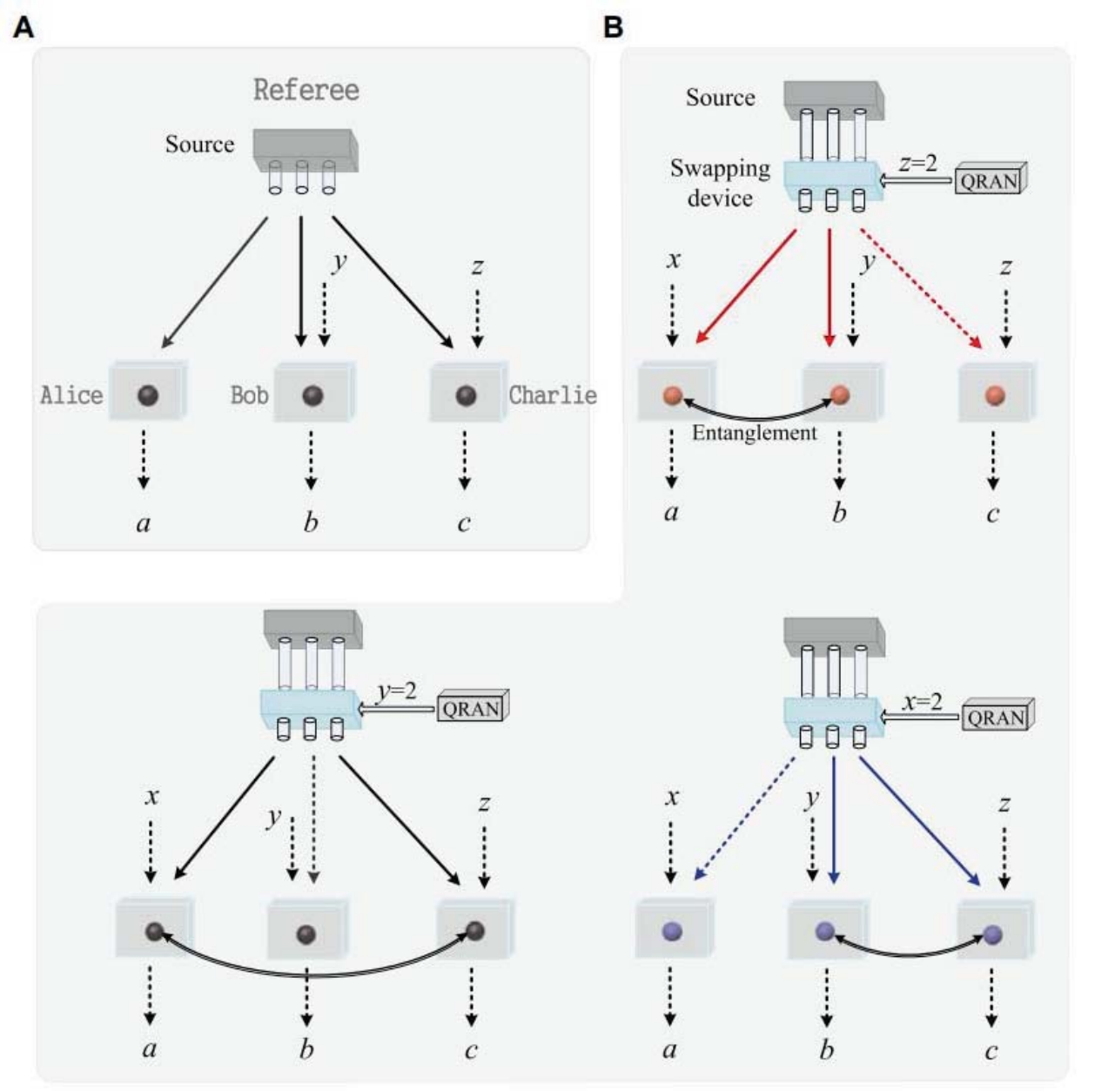}
\end{center}
\caption{\small \textbf{(Color online) Schematically tripartite Bell test}.}

{\small A. All observers share a variable $\lambda$ or any quantum state $\rho$ on Hilbert space $\mathcal{H}_A\otimes \mathcal{H}_B\otimes \mathcal{H}_C$. B. Each pair of two observers receive a bipartite state $\rho$ conditional on inputs $x,y,z\in \{0,1\}$. Each player receives a state for the input $x,y,z=2$. There are two outputs $a,b,c\in \{-1,1\}$. All the measurement devices can be regarded as black boxes. }
\label{fig-1}
\end{figure}

In this paper, we propose a unified method to construct multipartite monogamy relations from any linear Bell inequalities that are useful to witness general entanglement. We compare these Bell experiments in static and dynamical manners, as shown in Figure \ref{fig-1}. We show that when all observers share any global quantum entanglement or post-quantum resources statically, their measurement statistics have no quantum nonlocality beyond classical scenarios \cite{Fei,PB,RH}. We further demonstrate that by dynamically distributing a small-size entangled state, their measurement statistics show quantum nonlocality by violating the present multipartite monogamy relations. We implement a tripartite experiment using two-photon entangled states. This allows implementing a measurement-device independent witness of Bell nonlocality beyond previous Bell nonlocality experiments \cite{Freedman,Weihs,Kwiat, Zhang,Zhong, Vertesi,Rowe,Big,Storz2023,Ansmann,Mao,Xue,Wu}.

\section*{\Large Results}

\section*{Dynamical verification of Bell nonlocalities}

Consider a standard Bell test with two parties Alice ($\sA$) and Bob ($\sB$). Let $x$ and $y$ be the measurement settings of Alice and Bob respectively, and $a$ and $b$ for their outcomes. The joint probability of all the outputs $a$ and $b$ conditional on the inputs $x$ and $y$ allows the following decomposition \cite{Bell}:
\begin{eqnarray}
 P(a,b|x,y)=\int p(a|x,\lambda)p(b|y,\lambda)d\mu(\lambda),
\end{eqnarray}
where the conditional probabilities $p(a|x,\lambda)$ and $p(b|y,\lambda)$ depend on their input and the shared variable $\lambda$ with the probability distribution $\mu(\lambda)$. A general Bell inequality with correlations of local outcomes observed is defined as
\begin{eqnarray}
\label{bell}
\mathcal{L}(\sA,\sB)
\equiv \sum_{x,y} \sum_{a,b} \alpha_{x,y}^{a,b} P(a,b|x,y) \leq c,
\end{eqnarray}
where parameters $\alpha_{x,y}^{a,b}$ can be chosen positive and $c$ is the classical bound in hidden variable model \cite{Bell}.

In quantum scenarios, two parties share an entangled state $\rho$ on Hilbert space $\cH_A\otimes \cH_B$. Both parties perform positive-operator valued measurements (POVMs). Alice uses the measurement operators $\{M_{a|x}\}$ satisfying $M_{a|x}\geq 0$ and $\sum_{a}M_{a|x}=\mathbbm{1}$ with the identity operator $\mathbbm{1}$, and Bob uses $\{M_{b|y}\}$ satisfying $M_{b|y}\geq 0$ and $\sum_{b}M_{b|y}=\mathbbm{1}$. From the Born's rule, the joint probability of outputs is defined as
\begin{eqnarray}
P_Q(a,b|x,y)={\rm{}Tr}(M_{a|x}\otimes M_{b|y}\rho).
\end{eqnarray}
Violating the inequality (\ref{bell}) by the quantum statistics witnesses the nonlocality of the shared entanglement.
Now, consider an $n$-partite Bell experiment with separated parties, $\sA_1,\cdots, \sA_n$. For any two parties $\sA_i$ and $\sA_j$ they can verify the bipartite nonlocality from their shared bipartite entanglement violating the Bell's inequality (\ref{bell}) with their measurement statistics. But remarkably, these kinds of violations cannot be simultaneously realized even if all the parties $\sA_i$'s share an $n$-partite entanglement. Especially, we have the following monogamy relation
\begin{eqnarray}
\label{sig} \sum_{1\leq i<j\leq n} {\cal
L}(\sA_i,\sA_j)\leq \frac{1}{2}n (n-1).
\end{eqnarray}
where all the correlations of ${\cal
L}(\sA_i,\sA_j)$ are included in one multipartite Bell experiment. This monogamy relation can be proved by using the inequality \cite{PB}: $\sum_{i=2}^{n} {\cal
L}(\sA_1,\sA_i)\leq (n-1)c$.
This inequality holds for all the parties share post-quantum sources in terms of any every non-signaling theory. This means that in any standard multipartite Bell test any local bipartite correlations shows a total similar behaviours regardless of their shared resources. Instead, the violation of the inequality (\ref{sig}) will rule out the non-signaling principle.

The present monogamy of Bell inequalities can be further extended to characterize more than two parties. Let $P(a_1,\cdots, a_m|x_1,\cdots, x_m)$
 be the joint probability that all parties $\sA_1, \cdots, \sA_m$ observe the outcomes $a_1,\cdots, a_m$ conditional on the inputs $x_1,\cdots, x_m$, respectively. Consider any linear Bell inequality
\begin{eqnarray}
\label{bellm}
\mathcal{L}(\sA_1,\cdots, \sA_m)
\equiv \sum_{x_1,\cdots, x_m} \sum_{a_1,\cdots, a_m} \alpha_{x_1,\cdots, x_m}^{a_1,\cdots, a_m} P(a_1,\cdots,a_m|x_1,\cdots, x_m) \leq c',
\end{eqnarray}
which holds in the local hidden variable model \cite{Bell} or biseparable model \cite{Sy}, and $c'$ is a constant. Inspired by recent result \cite{PB} we have the following monogamy relation as
\begin{eqnarray}
\label{sigm}
\sum_{1\leq i_1\leq \cdots\leq  i_m\leq n} {\cal
L}(\sA_{i_1},\cdots, \sA_{i_m})\leq cC^{n}_m,
\end{eqnarray}
where the constant $C_m^n$ denotes the combination number, i.e., $C_m^n=n!/(n-m)!m!$. Although violating the inequality (\ref{bellm}) witnesses the Bell nonlocality of specific multipartite entanglement,  simultaneous violations are not allowed for any genuine multipartite entanglement. This demonstrates the trade-off among strengths of violations of any linear Bell inequalities under the non-signaling principle.

Now, we present a unified way to violate the monogamy inequality (\ref{sigm}) by using dynamical implementations while still adhering to the non-signaling principle. To show main idea, we consider a tripartite Bell test $\mathcal{G}_3$ shown in Figure \ref{fig-1}. There are three parties involved in this test: Alice ($\sA$), Bob  ($\sB$), and Charlie  ($\sC$). They will utilize a shared source to complete the test. Each party possesses three measurement settings denoted as $x(y, z) \in {0, 1, 2}$. Additionally, they have two outputs, denoted as $a(b, c) \in {0, 1}$, for each measurement setting, respectively.
Suppose three parties share a classical measurable variable. Inspired by the CHSH inequality  \cite{CHSH}, we obtain the following Bell inequality as
\begin{eqnarray}
\cB(\sA,\sB,\sC)&:=&A_0 B_0 C_2 +A_0 B_1 C_2+A_1 B_0 C_2
\nonumber
\\
&&-A_1 B_1 C_2+A_0 B_2 C_0+A_0 B_2 C_1\nonumber
\\
&&+A_1 B_2C_0 -A_1B_2C_1+A_2 B_0 C_0
\nonumber
\\
&& +A_2 B_0 C_1+A_2B_1 C_0-A_2B_1 C_1
\leq 6
\label{lineareq}
\end{eqnarray}
where $A_i,B_j,C_k$ are measurement observables with two outcomes $\{-1,1\}$. This can be proved by the fact that for each set of inputs $x, y, z$, the measurement statistic will violate the CHSH inequality if any two parties who have the inputs less than 2 win the CHSH game \cite{CHSH}, see the proof in SI \cite{SI}. This means the inequality (\ref{lineareq})  holds when all the parties share classical variables, quantum states, or post-quantum resources because of the monogamous features of nonlocalities \cite{RH,PB,Fei}. This means in the present setting, all the generated correlations consist of the same set regardless of the their shared resources. Moreover, this inequality holds for all the involved parties. This provides a form of global monogamy relationships going beyond all the previous results with local monogamy relationships \cite{RH,PB,Fei,Toner2009,RH,Regula2014}.

In quantum scenarios, each party in the tripartite test $\cG_3$ receives one particle from the source. Since the observers do not have prior knowledge about the state of the received particle, we impose a special restriction on each party: their local quantum measurement devices are black-boxes. This means that each party can only choose a local measurement box based on the input state, but they cannot modify the internal workings of their measurement devices. Under this assumption, we consider the present scheme as a measurement device-independent implementation of Bell test. There are three cases as follows:

Firstly, assuming that the source distributes a fixed three-particle state $\rho_{ABC}$ on Hilbert space $\cH_A\otimes \cH_B\otimes \cH_C$ to all parties, where Alice receives particle A, Bob receives particle B, and Charlie receives particle C. Similar to the standard Bell test this case depends on the independent identical  assumption of shared quantum states. The players can exploit any quantum correlations present in the shared state. However, all the quantum correlations satisfy the inequality (\ref{lineareq}), i.e., there is no quantum advantage beyond classical strategies.

Secondly, the source sends the particles passing through a swapping device. This can be regarded as a tripartite nonlocal game where the referee sends the swapped states to all observers conditional on their questions. The swapping operation $S=\sum_{i,j}|ij\rangle\langle ji|$ exchanges the states of two particles in the given state. For the input $z=2$, the source generates the state $\rho_{ABC}$ and sends it to all observers. For the input $y=2$, the source performs a swapping operation on particles B and C to get the state $\rho^{_{(y)}}_{ABC}\equiv S_{B,C} \rho S_{B,C}^\dag$ and then sends it to all observers. For the input $x=2$, the source performs swapping operation on particles A and C to get the state $\rho^{_{(x)}}_{ABC}\equiv S_{A,C} \rho S^\dag_{A,C}$ and sends it out. Suppose the source dynamically distributes a separable state $\rho_{ABC}=|\Phi\rangle\langle \Phi|_{AB}\otimes \rho_C$, where the generalized EPR state $|\Phi\rangle$ is defined by $|\Phi\rangle=\cos\theta|00\rangle+\sin\theta|11\rangle$ with $\theta\in (0,\frac{\pi}{4})$, and $\rho_C$ is a single state of particle C. This allows all observers simultaneously witness the Bell nonlocality of $|\Phi\rangle$ by violating the monogamy inequality (\ref{lineareq}) (see the proof in SI \cite{SI}) as
\begin{eqnarray}
\cB_{epr}>\cB_q=\cB_c=6
\label{SEa}
\end{eqnarray}
for any $\theta$ satisfying $\sin2\theta> \frac{2}{3}\sqrt{\frac{2}{5}}-\frac{1}{3}\approx 0.0866$. Here, the optimal winning probability is consistent with the maximal violation of the inequality (\ref{lineareq}) with the maximally entangled EPR states. This shows that the dynamical distribution of the EPR state allows all observers to achieve a quantum advantage beyond the static experiments with a fixed classical resource or any tripartite entangled states \cite{Sy}. The noise robustness of Werner state \cite{Werner} is easily followed for the linear inequality (\ref{lineareq}). We extend the analysis to multipartite Bell tests based the CHSH inequality or general Bell inequalities in SI \cite{SI}. This provides the first Bell experiment by combining quantum resources with dynamical resource distributing strategies beyond any others with single utilization of classical resources, quantum resources, or post-quantum resources.

\begin{figure}
\begin{center}
\includegraphics[scale =0.9]{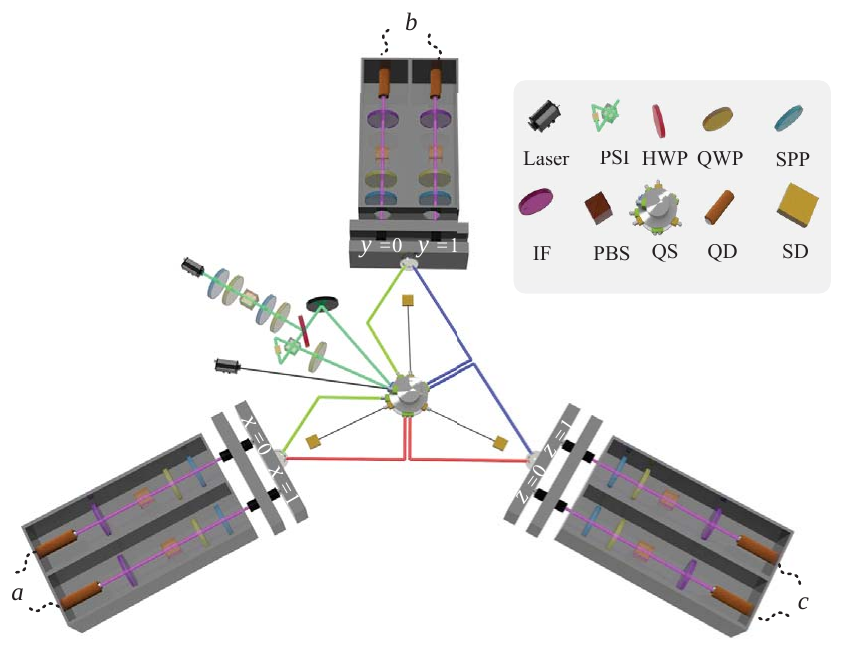}
\end{center}
\caption{\textbf{Sketch of the experimental setup.}}

{\small The SPDC source distributes maximally entangled two-photon pairs to build a probabilistic triangle quantum network. At one time, each player receives one photon and performs single-qubit measurements in Eq.(\ref{EPRn}). The SPDC source is based on a PPKTP crystal and is pumped by an ultraviolet pump pulse (405-nm, 80-MHz, 140-fs). Here, key elements include polarization Sagnac interferometer (PSI), half-wave plate (HWP), quarter-wave plate (QWP), splitting photon plate (SPP), interference filter (IF), polarizing beam splitter (PBS), quantum switch (QS) based on quantum random number generator (QRNG),  quantum detector (QD), and single photon detector (SD). }
\label{setup}
\end{figure}

\section*{Experimental results}

\begin{figure}
\begin{center}
\includegraphics[scale=0.75]{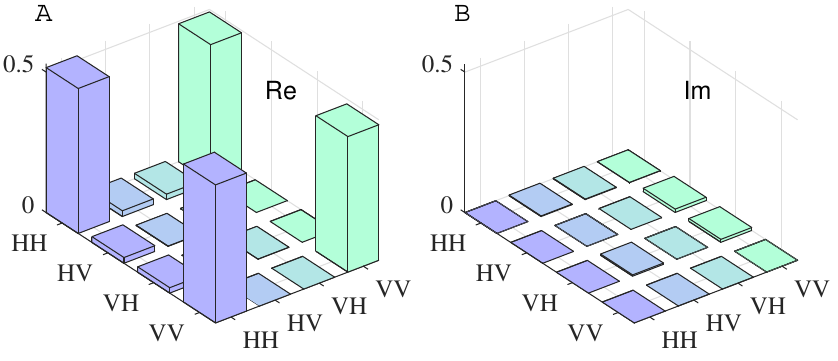}
\end{center}
    \caption{\textbf{Tomographic results of the EPR source.}}

{\small The left subplot represents the real part, while the right subplot represents the imaginary part of the reconstructed density matrices.}
    \label{source}
\end{figure}

To verify the Bell nonlocality, we first prepare a maximally entangled two-photon entangled state $|\psi\rangle=(|HH\rangle+|VV\rangle)/\sqrt{2}$, as shown in Figure ~\ref{setup}. We exploit the spontaneous parametric down-conversion (SPDC) source to generate polarimetric entangled photons \cite{Pan2012}. We encode horizontal (H) and vertical (V) polarization states as qubits $|0\rangle$ and  $|1\rangle$, respectively. The reconstructed density matrix is illustrated in Figure ~\ref{source}. The fidelity of the state $F=\langle \phi^+|\rho_{exp}|\phi^+\rangle$ is calculated to be $0.9905 \pm 0.0004$. The standard deviation is estimated assuming Poisson statistics. The $P$ value is $10^{-12}$ in our experiment.

\begin{figure}
    \begin{center}
\includegraphics[scale=0.65]{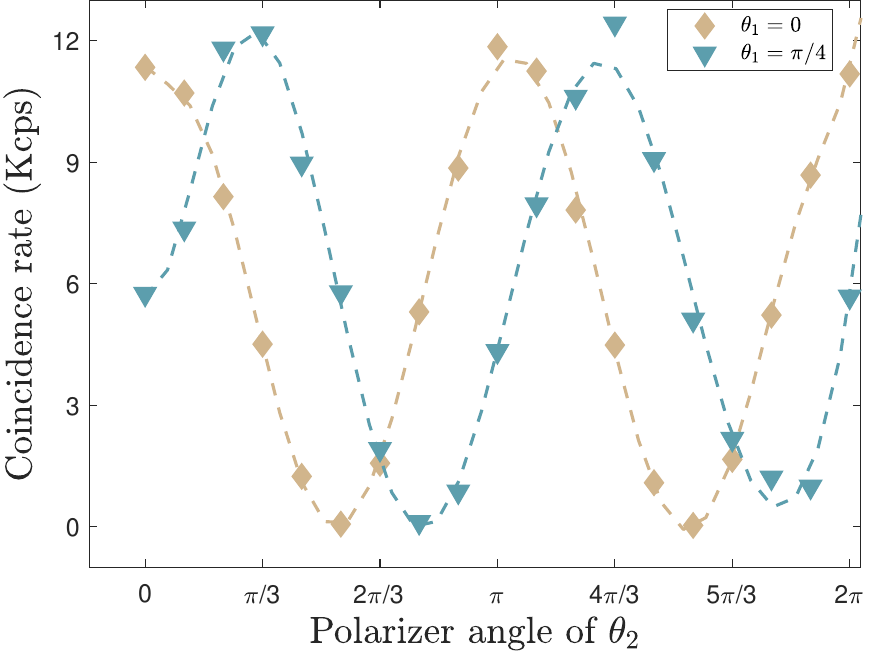}
\end{center}
    \caption{\textbf{Experimental visibilities of $\theta_1$ against the polarizer angle of $\theta_2$.}}

{\small The gray region corresponds to $\theta_1$ being fixed at an angle of 0, while the blue region corresponds to $\theta_1$ being fixed at an angle of $\pi/4$. Combining a HWP and a PBS in front of the detectors acts as a polarizer box, where the polarizer angle is given by $\theta^{\rm{pol}}=2\theta^{\rm{HWP}}$.}
    \label{interference}
\end{figure}

Figure ~\ref{interference} presents a graphical representation of the experimental coincidence rate behavior of our EPR source. The overlaid dotted curves are the best sinusoidal fits to the corresponding data. We achieve this by fixing the polarizer angle $\theta_1$ at 0 and $\pi/4$, while varying the polarizer angle $\theta_2$ to align with the theoretical prediction. We measure the experimental visibilities of $V_{HV}= 0.9956\pm 0.0009$ and $V_{DA}=0.9818\pm 0.0019$. The standard deviation is estimated according to Poisson statistics. These experimental visibilities exhibit a substantial and consistent deviation from the predictions of classical physics which predicts that $V\leq 0.71$.

During the experiment, we let the QRNG dynamically distribute the maximally entangled EPR state $|\Phi\rangle= (|HH\rangle+|VV\rangle)/\sqrt{2}$ to two out of the three parties, Alice, Bob, and Charlie. QRNG generates a set of sequences (consisting of 0, 1, and 2) according to uniform random distribution, where 0 encodes the case that the entangled source will be sent to Alice and Bob with the experimental probability, 1 encodes the case for Bob and Charlie and 2 encodes the case for Alice and Charlie. The experimental coincidence probabilities are shown as $p_{0}=0.33549$, $p_{1}=0.33241$, and $p_{2}=0.33210$.

In our experiment, we record the basis choices $(x, y, z)$ and the corresponding readout result $(a,b,c)$ for each of the $n$ trials. Utilizing these recorded values, we calculate the correlator $\langle A_{x}B_yC_{z}\rangle$ for all possible combinations of measurement bases. In each of the twelve consecutive experiments, we took 400s to complete the evaluation. We perform a post-selection process, retaining 16 sets of coincidences, and present the corresponding experimental values in Table \ref{correlators}. The coincidence rates are shown in Tables S1-S3 in SI \cite{SI}.  Notably, the experimental measurement (\ref{lineareq}) yields a value of 7.5348, which violates the maximal bound stipulated in the inequality (\ref{lineareq}). In our experiment, we acknowledge the presence of two common loopholes that are frequently encountered in nonlocality experiments: the locality loophole and the postselection loophole \cite{AKAZ}.

\begin{table}\normalsize
    \centering
    \caption{Experimental correlators $\langle A_{x}B_yC_z\rangle$ for twelve combinations of basis choices $(x, y, z)$.}
  \begin{tabular}{c|c}
  \hline\hline
     Correlator & Experimental values
    \\ \hline
     $\langle A_0 B_0 C_2 \rangle$    &  $0.6856\pm 0.0062$ \\
     $\langle A_0 B_1 C_2 \rangle$    &
     $0.6972\pm 0.0060$ \\
     $\langle A_1 B_0 C_2\rangle$ & $0.6794\pm 0.0063$
     \\
     $\langle A_1 B_1 C_2\rangle$ &  $-0.6698\pm 0.0064$
    \\ \hline
  $\langle A_0 B_2 C_0\rangle$ & $0.9710\pm 0.0012$
  \\
  $\langle A_0 B_2 C_1\rangle$ & $0.0215\pm 0.0129$
  \\
  $\langle A_1 B_2C_0\rangle$ & $0.0190\pm 0.0130$
  \\
  $\langle A_1B_2C_1\rangle$ & $0.9674\pm 0.0015$
  \\ \hline
  $\langle A_2 B_0 C_0\rangle$ & $0.7012\pm 0.0060$
  \\
  $\langle A_2 B_0 C_1\rangle$ & $0.6823\pm 0.0061$
  \\
  $\langle A_2B_1 C_0\rangle$ & $0.6737\pm 0.0062$
  \\
  $\langle A_2B_1 C_1\rangle$ & $-0.6716\pm 0.0062$ \\
 \hline\hline
    \end{tabular}
    \label{correlators}
\end{table}

\section*{\Large Discussion}

We have presented tripartite monogamy relationships of violating CHSH inequality \cite{CHSH}, which can be regarded as a wired Bell inequalities. This kind of Bell inequalities provide the trade-off among strengths of violations of partial Bell inequalities under only the non-signaling condition \cite{Fei,PB,RH}. We have further shown that there are quantum violations when a source generates a flying EPR state \cite{EPR}. Although the shared state has changed during in each round of experiment, it is compatible with the independent and identical distribution assumptions under specific input distribution. The swapping operation being performed by the source can be regarded as a post-selection strategy in a standard Bell experiments with three generated states \cite{Bell,BCPS}. This provides a witness of Bell nonlocality beyond previous experiments \cite{Freedman,Weihs,Kwiat,Zhang,Zhong,Vertesi,Rowe,Big,Storz2023}.  If we regard the present Bell tests as computation tasks that are distributed across a network of interconnected nodes dynamically adapting to changing conditions, the present quantum advantage means that combining quantum resources with dynamical resource strategies are going beyond any other single utilization of resources. This  harnesses the collective computational power of distributed systems to dynamical changes and effectively tackle large-scale problems, and provides a new insight in designing quantum distributed tasks or quantum cryptography on dynamical quantum Internet \cite{Kim}. This also intrigues to explore the inherent nonclassicality of general multipartite Bell tests. We present an extension with any linear Bell inequality in SI.

In conclusion, we have presented a unified way to construct monogamy relationships using any Bell inequality. Violating these monogamous relationships is distinguished for all parties sharing general sources. The present model is useful for witnessing any entangled pure state by lifting it into a larger space. This can be regarded as a feature of entangled states combined with implementation strategies beyond previous static quantum entanglement experiments. Our results are interesting in Bell theory, theoretical computer science, and distributive computations.

\section*{\Large Experimental procedures}

\section*{\small Resource Availability}

\subsubsection*{Lead Contact}

Further information and requests for resources should be directed to the lead contact Ming-Xing Luo (mxluo@swjtu.edu.cn).

\subsubsection*{Materials available}

This paper did not generate new materials.

\subsubsection*{Data and code availability}

The experimental data are available from the corresponding author upon reasonable request. This paper did not report original code.

\section*{Experimental setup}

We build the dynamical implementation of tripartite nonlocal game $\cG_3$ with a biseparable state by a spontaneous parametric down-conversion (SPDC) source, which produces polarisation entangled photon pairs to supply the entangled photonic link. The experimental setup is illustrated in Figure \ref{setup}. To create a probabilistic triangle network, a fixed EPR source distributes pairs of two photons to Alice, Bob, and Charlie. The allocation of these photon pairs to the parties is determined by using the QS based on QRNG with uniform distribution.

The source of EPR states is generated by a cavity-stabilized Ti:sapphire pump laser operating at a central wavelength of 405 nm. This laser generates signal and idler photons with wavelengths symmetrically distributed around 810 nm through the process of SPDC. The SPDC source comprises a 15*2*1 mm periodically poled KTiOPO$_4$ (PPKTP) crystal, which is cut in beam-like type-II phase matching configurations, providing advantages in terms of high entangled photon brightness (0.34 MHz) and efficient light collection ($60\%$).

To generate polarization-entangled photon pairs, we showcase a variation of the bidirectionally pumped down-conversion source that utilizes a PSI. The pump pulse is divided into two beams (horizontally and vertically polarized pump) with equal power using QWPs, HWPs, a PBS, and a dual-wavelength polarizing beam splitter. Each of these beams, with a pump power of 18 mW, is used to stimulate the SPDC source to generate EPR states. A pump photon has an equal chance of undergoing downconversion in the PPKTP crystal, resulting in photon pairs with the polarization state $|H\rangle_1|V\rangle_2$. If the pair is generated by the horizontally polarized pump component, it goes through a dual-wavelength half-wave plate, which causes a rotation to the state $|V\rangle_1|H\rangle_2$. After passing through the PSI, the second particle of the output state undergoes a polarization rotation of $\pi/2$ at the HWP. After compensating for spatial and temporal factors, the final state of the photon pair is given by $(|H\rangle_1|H\rangle_2 + |V\rangle_1|V\rangle_2)/\sqrt{2}$.

The qubits are encoded in the polarisation degree of freedom of the photons. Following the rules of nonlocal game in SI, different inputs are associated with different CHSH operators. The measurement device at each party is a polarisation analysis system which consists of a QWP, an HWP, a PBS, an IF, and a QD.

\section*{Experimental method}

Each party's detector has three settings $x,y,z\in \{0,1,2\}$ with three outcomes for each setting. For two measurement settings $\{0,1\}$, the measurement bases corresponding to the detector settings of each party are defined as
\begin{eqnarray}
&&A_{0}=C_0=\sigma_z,  A_{1}=C_1=\sigma_x, B_{y}=\frac{1}{\sqrt{2}}(\sigma_z+(-1)^y\sigma_x).
\label{EPRn}
\end{eqnarray}
For the measurement setting $x_i=3$, each party will output a bit based on a classical random number generator. Here, the quantum measurement device for each party is regarded as a black box which is controlled by a quantum switcher. Especially, for a given input $x_i$, the quantum switcher will transfer the state into the corresponding input channel of the measurement device.

For the setting $z=2$, we evaluate the tripartite correlation $\langle A_{x}B_{y}C_{z}\rangle=\langle A_{x} B_{y}\rangle \langle C_{z}\rangle $. This allows us to evaluate the experimental correlators using the photonic coincidence probabilities of measurements made locally by each pair of two quantum observers on their respective subsystems. In our case, one is the measurement outcomes on the signal and idler photons from the SPDC source, and the other is from a classical random signal. Especially, we compute the tripartite correlation as
\begin{eqnarray}
\langle A_{x}B_{y}C_{z}\rangle=
\frac{\sum_{a,b=\pm 1}(-1)^{s(a,b)}N^{a,b}_{x,y}}{\sum_{a,b=\pm 1 }N^{a,b}_{x,y}}
\times
\frac{\hat{N}^{+}_{2}-\hat{N}^{-}_{2}}{\hat{N}^{+}_{2}+\hat{N}^{-}_{2}}
\end{eqnarray}
where $N_{x,y}^{a,b}$ denote photon coincidences, $s(a,b)$ denotes a function equal to $1$ if there are odd number of $-1$, and 2 otherwise, and $\hat{N}^{\pm}_{2}$ denotes the coincidence of classical signal which is evaluated by using random number generator. Similar evaluations are for $x=2$ or $y=2$. With these correlations, we then calculate the measurement operator (\ref{lineareq}) and winning probability.

Early Bell test experiments utilized the standard deviation as a metric to gauge the statistical significance of a Bell inequality violation \cite{Aspect1981}. This approach is constrained by two assumptions: the Gaussian distribution assumption and the independence of each trial result. To address these limitations, we adopt a statistical analysis that relies on the evaluation of the $P$ value \cite{HTP}.

\section*{Acknowledgements}

This work was supported by the National Natural Science Foundation of China (Nos.62172341, 61772437, 12075159), Sichuan Natural Science Foundation (No.2023JQ00447), Academy for Multidisciplinary Studies, Capital Normal University, and the Academician Innovation Platform of Hainan Province.

\section*{Author Contributions}

Y.H. and M.X. conceived the study. Y.H. and M.X. designed the experiment. Y.H. and X.Z. conducted the experiments. All authors wrote and reviewed the paper.

\section*{Declaration of Interests}

The authors declare no competing interests.

\end{document}